# Thomson and Collisional Regimes of In-Phase Coherent Microwave Scattering Off Gaseous Microplasmas

Adam R. Patel[1], Apoorv Ranjan[1], Xingxing Wang[1], Mikhail N. Slipchenko[2], Mikhail N. Shneider[3], and Alexey Shashurin[1*]

[1] School of Aeronautics and Astronautics, Purdue University, West Lafayette, IN, USA

[2] School of Mechanical Engineering, Purdue University, West Lafayette, IN, USA

[3] Mechanical and Aerospace Engineering, Princeton University, Princeton, NJ, USA

## Abstract

The total number of electrons in a classical microplasma can be non-intrusively measured through elastic in-phase coherent microwave scattering (CMS). Here, we establish a theoretical basis for the CMS diagnostic technique with an emphasis on Thomson and collisional scattering in short, thin unmagnetized plasma media. Experimental validation of the diagnostic is subsequently performed via linearly polarized, variable frequency microwave scattering off laser induced air-based microplasmas with diverse ionization and collisional features. Namely, conducted studies include a verification of short-dipole-like radiation behavior, plasma volume imaging via intensified charge-coupled device (ICCD) photography, and measurements of relative phases, total scattering cross sections, and total number of electrons $N_e$ in the generated plasma filaments following absolute calibration using a dielectric scattering sample. Findings of the paper suggest an ideality of the diagnostic in the Thomson "free-electron" regime – where a detailed knowledge of plasma and collisional properties (which are often difficult to accurately characterize due to the potential influence of inhomogeneities, local temperatures and densities, present species, and so on) is unnecessary to extract $N_e$ from the scattered signal.

---

*Corresponding Author, ashashur@purdue.edu

# Introduction

The scattering of electromagnetic radiation by an ionized gas is coupled with a multitude of plasma parameters – establishing a fundamental pathway for non-intrusive diagnostics with high spatial and temporal resolution [1]. In a manifestation devised by Shneider and Miles [2], the total number of electrons ($N_e$) present in a collisional, weakly-ionized, and unmagnetized small plasma object can be measured through the constructive coherent elastic scattering of microwaves. By irradiating the object with a linearly polarized, uniformly distributed long-wavelength microwave field (for a plasma column of length $L$ and diameter $D$, the skin depth $\delta \gg D$ and microwave wavelength $\lambda \gg D, L$), the plasma is periodically polarized and consequently emits short-dipole-like (short relative to the wavelength) radiation. The scattering is thus deemed in-phase coherent for the considered case of minimal phase variation amongst wavelets (scattered electric field contributions from individual electrons) at a far-field detecting location. Emitted radiation can then be measured/isolated using temporally resolved frequency mixer technologies and attributed to an absolute electron count after system calibration [3]. Both calibrated and uncalibrated variants of in-phase coherent microwave scattering (CMS) have become valuable non-intrusive diagnostic techniques in applications ranging from photoionization and electron-loss rate measurements [3] [4] [5] [6] [7] [8] [9] [10] to trace species detection [11], gaseous mixture and reaction characterization [12] [13] [14], molecular spectroscopy [15], and standoff measurement of local vector magnetic fields in gases through magnetically-induced depolarization [16]. Notable advantages of the technique include a high sensitivity, good temporal resolution, low shot noise [11], non-intrusive probing, and the capability of time gating due to continuous scanning.

A careful understanding of electromagnetic-plasma and scattering interactions is essential for the proper extraction of information on total electron numbers and subsequent derivable quantities (e.g., the electron number density $n_e$) from CMS. Vast historical art exists for weakly-ionized, strongly collisional microplasmas in which electron motion is restricted by electron-neutral collisions [4] [5] [9] [10]. However, application of the diagnostic to collisionless plasmas has remained relatively unexplored (no $N_e$ correspondence is given by [11] [13] [17]) – despite garnering some attention in recent years [18] due to a prevalence in studies on electric propulsion devices [19], photoionization in low-pressure conditions [20] [21], etc. Such plasmas are particularly advantageous for CMS in that a detailed knowledge of plasma and collisional properties (often susceptible to inhomogeneities, local temperatures and densities, present species, and so on) may not be required to characterize scattering.

Here, we establish a theoretical basis for coherent microwave scattering in the Thomson "free-electron" regime at reduced gas pressures and its transition to the conventional collisional scattering regime at atmospheric pressures – encompassing classical small plasma objects with diverse ionization and collisional properties. We subsequently present an experimental verification of this basis via variable frequency microwave scattering off laser-ionized air-based filaments for a range of pressures 1-760 Torr.

## Results and Discussion

**Theory: Coherent plasma scattering of microwaves.** To establish a physical intuition for CMS, we first consider scattering contributions at a distance $R \gg V^{1/3}$ ($V^{1/3}$ is a characteristic scale of the scattering volume $V$) from $N$ nonrelativistic electron scatterers assuming each electron (distinguished by indices $j, l$) emits an electric field amplitude $E_{S,0}$ at the observation location with frequency $\omega$ and relative phase $\Phi_j$. Scattering of microwave radiation from relatively massive ions will be assumed negligible in the context of this work. The time-averaged scattered power per unit area at detector location $\langle dP_S/dA \rangle$ in SI units is subsequently given in Eq. 1, in which $c$ is the speed of light in vacuum and $\varepsilon_0$ is the vacuum dielectric permittivity [1]:

$$\langle \frac{dP_S}{dA} \rangle = c\varepsilon_0 \langle \sum_{j=1}^{N} \boldsymbol{E}_{jS} \cdot \sum_{l=1}^{N} \boldsymbol{E}_{lS} \rangle = \frac{c\varepsilon_0}{2} N E_{S,0}^2 + c\varepsilon_0 \langle \sum_{\substack{j,l=1 \\ l \neq j}}^{N} \boldsymbol{E}_{jS} \cdot \boldsymbol{E}_{lS} \rangle \quad \text{Eq. 1}$$

where the first and second terms in Eq. 1 reflect noncollective and collective scattering phenomena, respectively. The collective scattering term is alternatively phrased as the "coherent" expression and can yield various contributions to the total scattered power as outlined below. First, in the case of full constructive interference between wavelets (deemed in-phase coherent), the collective term yields $c\varepsilon_0 \langle \sum_{\substack{j,l=1 \\ l \neq j}}^{N} \boldsymbol{E}_{jS} \cdot \boldsymbol{E}_{lS} \rangle = \frac{c\varepsilon_0}{2} N(N-1) E_{S,0}^2$ and thus a maximum in the scattered power per unit area is reached $\langle \frac{dP_S}{dA} \rangle_{max} = \frac{c\varepsilon_0}{2} N^2 E_{S,0}^2$. Second, for phase-locked electrons exhibiting certain phase shifts at the detector location, the contribution of the collective term to the total power would be smaller $\langle \frac{dP_S}{dA} \rangle < \langle \frac{dP_S}{dA} \rangle_{max}$. The summation term is consequently zero $\langle \sum_{\substack{j,l=1 \\ l \neq j}}^{N} \boldsymbol{E}_{jS} \cdot \boldsymbol{E}_{lS} \rangle = 0$ for phase-locked electrons exhibiting uniform and continuous relative phase distributions of $\Phi_j$ on the interval $[0, 2\pi)$. Third, if we are to consider significant phase randomization over the temporal resolution of the detecting equipment (i.e., from thermal motion, Doppler broadening, etc.) [22], each term in the summation yields zero $\langle \boldsymbol{E}_{jS} \cdot \boldsymbol{E}_{lS} \rangle_{l \neq j} = 0$. This case refers to the incoherent scattering regime and, from Eq. 1, the time-averaged scattered power per unit solid angle is subsequently evaluated as $\langle \frac{dP_S}{dA} \rangle = \frac{c\varepsilon_0}{2} N E_{S,0}^2$.

The position, velocity, and acceleration of individual scatterers in an ensemble (with respect to the detector) are central factors in determining coherency [1]. For Thomson scattering of electromagnetic waves on free plasma electrons at optical frequencies, the following simplified criteria is commonly utilized. If the parameter $\alpha = \frac{1}{k\lambda_D} \ll 1$ (where $k = |\boldsymbol{k}_I - \boldsymbol{k}_S|$ and $\boldsymbol{k}_I, \boldsymbol{k}_S$ are wavevectors of the incident and scattered radiation and $\lambda_D$ is the Debye radius), then the wave resolves the motion of individual uncorrelated free electrons within the Debye radius and the scattering is incoherent. Otherwise, for $\alpha \gg 1$, the waves "see" the collective motion of different plasma elements and the scattering is coherent [1] [23] [24].

In this work, we consider irradiation with microwave frequencies – ensuring that scattering occurs in the coherent regime. This conclusion generally follows from the fact that the microwave wavelength greatly exceeds the optical wavelength by about 5 orders of magnitude. Therefore, the signals generated at the detector from each individual electron in the plasma volume are coherent (e.g., phase randomization due to thermal motion and associated electron displacements comparable to $\lambda$ occurs at a very slow time scale, Doppler broadening is unable to randomize phases for reasonable detector distances). More specifically, this work deals with a weakly-ionized plasma ellipsoid of length $L \leq 1$ cm and diameter $D \leq 0.5$ mm – where the linearly polarized microwave field ($D \ll L < \lambda \sim 3$ cm) is oriented along the ellipsoid length and the skin depth $\delta \gg D$. Therefore, all plasma electrons "see" the same intensity and phase of the incident microwave field. Phase differences generated by electrons from different portions of the plasma volume are then negligible as the considered experimental detector distance $R = 10$ cm greatly exceeds $\frac{L^2}{\lambda} \sim 0.3$ cm. In this situation, full constructive interference between wavelets occurs and indicates that CMS scattering refers to the aforementioned in-phase coherent case.

**Theory: Model for microwave scattering from a short, thin plasma ellipsoid.** In this section, we consider scattering of linearly polarized electromagnetic radiation off a plasma ellipsoid with the major axis (of length $L$) oriented parallel to the incident field. The considered ellipsoid will be regarded as short ($L < \lambda$), homogeneous, and thin ($D \ll \lambda, \delta$). The corresponding radiation characteristics can be sufficiently described by short dipole theory [25] [26] and an accompanying plasma fluid model for the motion of plasma electrons. Generally, the equation of motion for a nonrelativistic electron in the swarm is:

$$m\ddot{\mathbf{s}} = -e(\mathbf{E} + \dot{\mathbf{s}} \times \mathbf{B}) - m\nu_m \dot{\mathbf{s}} \qquad \text{Eq. 2}$$

where $\mathbf{s}$ is the electron displacement vector, $e$ is the absolute electron charge, $m$ is the electron mass, $\mathbf{E}$ and $\mathbf{B}$ are local electric and magnetic fields (respectively), and the effective collisional frequency for momentum transfer $\nu_m = \sum_t \nu_{m,t}$ reflects the sum total of electron collisions with specie index $t$. The collisional frequency ($\nu_m = \nu_{m,eg} + \nu_{m,ei} = n_g \sigma_{tr} \text{v}_{Te} + n_e \sigma_{Coul} \text{v}_{Te}$, for which $\text{v}_{Te}$ is the electron thermal velocity and $\sigma$ are cross sections) is governed by electron-gas collisions $\nu_{m,e-g}$ for ionization degrees on the order of $\alpha < 10^{-2} - 10^{-3}$, while electron-ion collisions $\nu_{m,e-i}$ may become appreciable at higher ionization degrees [27]. $\nu_m$ resultingly exhibits complicated dependencies on temperatures, densities, and present species.

Eq. 2 is subsequently reduced to the following form noting a unidirectional electric field and assuming negligible **B**-field interactions (similar to approach used in the Lorentz oscillator model [28]):

$$\ddot{s} + \nu_m \dot{s} = -\frac{e}{m}E \qquad \text{Eq. 3}$$

Where the total electric field inside the plasma volume ($E$) is composed of an incident irradiating field ($E_I$) and a depolarizing field ($E_{dep}$) due to an accumulation of charges at the ellipsoid ends.

The problem can be restated in complex formalism for an $e^{i\omega t}$-process considering the incident electric field in the form $E_I = E_{I,0}\cos(\omega t) = \text{Re}\{\tilde{E}_{I,0}e^{i\omega t}\}$ and the electron displacement in the form $s = s_0\cos(\omega t + \Phi) = \text{Re}\{\tilde{s}_0 e^{i\omega t}\}$ to account for the phase-lag ($\Phi$) between the electron displacement and the incident electric field ($\tilde{E}_{I,0}$ and $\tilde{s}_0$ refer to as complex amplitudes). Then, since we examine scattering off a thin ($D \ll \lambda, \delta$), homogeneous plasma ellipsoid with the incident field oriented along its length, the established complex amplitude of the total electric field inside the plasma can be written as follows: $\tilde{E}_0 = \frac{\tilde{E}_{I,0}}{1+\xi(\tilde{\varepsilon}-1)}$ where $\tilde{\varepsilon} = \varepsilon' - i\varepsilon'' = \left(1 - \frac{\omega_p^2}{\omega^2+\nu_m^2}\right) - i\frac{\omega_p^2}{\omega^2+\nu_m^2}\frac{\nu_m}{\omega}$ is the complex dielectric permittivity for the plasma, $\omega_p = \sqrt{\frac{n_e e^2}{m\varepsilon_0}}$ is the plasma frequency, and $\xi$ is the depolarization factor for an ellipsoid [29] [30]. The established solution for the complex electron displacement amplitude is correspondingly written as $\tilde{s}_0 = \frac{-e\tilde{E}_{I,0}}{m(1+\xi(\varepsilon'-1)-i\xi\varepsilon'')}\frac{1}{(-\omega^2+i\omega\nu_m)} = \frac{-e}{m}\frac{\tilde{E}_{I,0}}{(\xi\omega_p^2-\omega^2)+i\nu_m\omega}$. Finally, the actual displacement amplitude $s_0$ and phase with respect to the incident field $\Phi$ is reduced to the form:

$$s_0 = |\tilde{s}_0| = \frac{e}{m}\frac{E_{I,0}}{\sqrt{\left(\xi\omega_p^2-\omega^2\right)^2 + (\nu_m\omega)^2}} \qquad \text{Eq. 4}$$

$$\tan(\Phi) = \frac{-\nu_m\omega}{\xi\omega_p^2-\omega^2} \qquad \text{Eq. 5}$$

Note that an identical result can be obtained by evaluating the depolarizing field as $E_{dep} = \frac{\xi m \omega_p^2}{e} s$ (or by including restoring force term $\xi \omega_p^2 s$ in the left side of Eq.3) [16].

The total dipole moment of the oscillating plasma channel can then be characterized by the amplitude $d_0 = es_0 \int n_e(r,z) 2\pi r dr dz = es_0 N_e$. In the far-field (short dipole criteria $R > 2\lambda$ [25] [31]), the total time-averaged scattered power can be treated as a point dipole and related to the dipole moment amplitude as $P_S = \frac{\ddot{d}^2}{6\pi\varepsilon_0 c^3} \rightarrow \langle P_S \rangle = \frac{\omega^4 d_0^2}{12\pi\varepsilon_0 c^3}$ and, using Eq. 4 and the incident microwave field intensity $I_I = \frac{1}{2}\varepsilon_0 c E_{I,0}^2$, finally reduced to the form:

$$\langle P_S \rangle = \frac{e^4}{6\pi m^2 \varepsilon_0^2 c^4} \frac{I_I \omega^4}{\left(\xi \omega_p^2 - \omega^2\right)^2 + (\nu_m \omega)^2} N_e^2 \qquad \text{Eq. 6}$$

With associated far-field total and differential cross sections for a point dipole:

$$\sigma_{Tot} = \frac{\langle P_S \rangle}{I_I} = \frac{e^4}{6\pi m^2 \varepsilon_0^2 c^4} \frac{N_e^2 \omega^4}{\left(\xi \omega_p^2 - \omega^2\right)^2 + (\nu_m \omega)^2} = \sigma_{Th} \frac{\omega^4}{\left(\xi \omega_p^2 - \omega^2\right)^2 + (\nu_m \omega)^2} N_e^2 = \sigma_e N_e^2 \qquad \text{Eq. 7}$$

$$\frac{d\sigma_{Tot}}{d\Omega} = \frac{3}{8\pi} \sigma_{Tot} \sin^2(\theta) \,;\, \frac{dP_S}{d\Omega} = \frac{d\sigma_{Tot}}{d\Omega} I_I \qquad \text{Eq. 8}$$

where $\sigma_{Th} = \frac{e^4}{6\pi m^2 \varepsilon_0^2 c^4}$ is the Thomson cross-section [32] [33] and $\sigma_e = \sigma_{Th} \frac{\omega^4}{\left(\xi \omega_p^2 - \omega^2\right)^2 + (\nu_m \omega)^2}$ is the total cross-section of an individual electron.

Three distinct coherent scattering regimes can be interpreted from Eq. 7 corresponding to the dominant denominator term – as detailed below and summarized in Table I. First, the Thomson scattering regime can be identified which is associated with free plasma electrons oscillating in-phase with the incident microwave field (the $\omega^4$ term is dominant in the denominator of Eq. 7). The total cross-section of an individual electron then coincides with the classical Thomson cross section $\sigma_e = \sigma_{Th}$ and is thus independent of the microwave wavelength $\lambda$. Second, the collisional scattering regime (often referred to as the quasi-Rayleigh regime) refers to collision dominated electron motion ($\nu_m^2 \omega^2$ term is dominant) with displacement oscillations shifted 90° with respect to the $E_I$-field oscillations. The individual electron cross section $\sigma_e = \sigma_{Th}\left(\frac{\omega}{\nu_m}\right)^2$ correspondingly scales as $\propto \frac{1}{\lambda^2}$. Previous experiments on microwave scattering off microplasmas have been conducted in this regime as discharges were operated at elevated pressures [3] [4] [5] [9] [10] [34]. A detailed knowledge of $\nu_m$ is consequently required for an accurate characterization of scattering. Third, the Rayleigh scattering regime can be observed which is associated with restoring-force-dominated electron motion ($\xi^2 \omega_p^4$ term is dominant) with $s$ oscillating in antiphase with $E_I$. The total individual electron cross section then evaluates as $\sigma_e = \sigma_{Th}\left(\frac{\omega^2}{\xi \omega_p^2}\right)^2$ and classically scales with the wavelength as $\propto \frac{1}{\lambda^4}$. Rayleigh scattering is not comprehensively explored in this work due to relatively low plasma densities and depolarization factors for the encountered experimental conditions (this regime can be somewhat exotic for practical implementation). However, the current work reports the first demonstration of in-phase coherent Thomson scattering of microwaves off miniature plasma objects.

A visualization of scattering regimes is illustrated in Fig. 1 for 11 GHz microwaves (as utilized in the experiment) using contour plots of individual electron scattering cross sections ($\sigma_e$) and phase-lags between $s$ and $E_I$ ($\Phi$). An air-based, singly ionized thin/short plasma ellipsoid with depolarization factor $\xi = 0.01$ and electron temperature $T_e = 0.9$ eV is considered for Fig. 1 – where the collisional frequency was subsequently calculated as $\nu_m = n_{O_2}\sigma_{eO_2}\text{v}_{Te} + n_{N_2}\sigma_{eN_2}\text{v}_{Te} + n_e\sigma_{Coul}\text{v}_{Te}$ (for values and assumptions on electron-neutral collisions, see Methods – Electron number measurements). The Coulomb cross section for electron-ion collisions was evaluated as $\sigma_{Coul} = \frac{4\pi}{9}\frac{e^4 \ln(\Lambda)}{(kT_e)^2}, \Lambda = \frac{3}{2\sqrt{\pi}}\frac{(kT_e)^{3/2}}{e^3 n_e^{1/3}}$ [27].

Table I: Coherent Microwave Scattering Regimes.

| Regime | Condition | Coherent scattering cross-section, $\sigma_e$ | Phase Shift, $\Phi$ |
|---|---|---|---|
| Thomson | $\omega^2 \gg \nu_m\omega, \xi\omega_p^2$ | $\sigma_{Th}$ | 0° |
| Collisional | $\nu_m\omega \gg \|\xi\omega_p^2 - \omega^2\|$ | $\sigma_{Th}\left(\dfrac{\omega}{\nu_m}\right)^2 \propto \dfrac{1}{\lambda^2}$ | 90° |
| Rayleigh | $\xi\omega_p^2 \gg \omega^2, \nu_m\omega$ | $\sigma_{Th}\left(\dfrac{\omega^2}{\xi\omega_p^2}\right)^2 \propto \dfrac{1}{\lambda^4}$ | 180° |

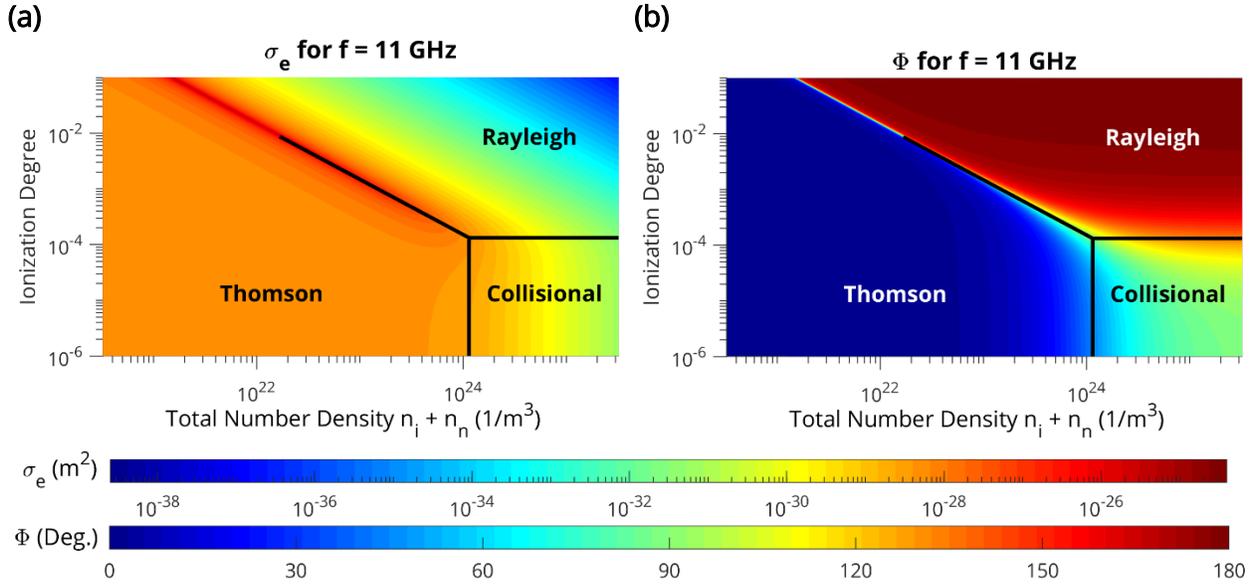

**Fig. 1 Scattering regimes.** Contour plots for f = 11 GHz (a) total individual electron scattering cross sections $\sigma_e$ – calculated from Eq. 7 and (b) phase-lag between electron displacement and the incident microwave field – calculated from Eq. 5. The corresponding scattering regimes (Thomson, collisional, and Rayleigh) are additionally depicted. An air-based, singly ionized short/thin plasma ellipsoid with electron temperature $T_e$ = 0.9 eV (relevant to considered experimental plasmas) and depolarization factor $\xi = 0.01$ is considered.

**Experimental results and discussions.** We will now experimentally demonstrate the Thomson CMS regime and utilize the one-dimensional model described above for measurements of electron numbers and electron number densities in the Thomson scattering regime. To this end, we first consider unmagnetized, laser induced $\leq$ 760 Torr weakly ionized air-based microplasmas predominantly realized via the O$_2$ resonance-enhanced multiphoton ionization (REMPI) mechanism (see Methods for details). An Abel inverted image (gate width 100 ns, 250 accumulations) of the produced non-equilibrium plasma for a background pressure of 760 Torr is illustrated in Fig. 2(a). One can see that the plasma can be regarded as an oblate spheroid with upper bound (due to imaging uncertainties and diffusion) semi-axis estimates of $\mathfrak{a} = \mathfrak{b} = 280$ μm and $\mathfrak{c} = 4.5$ mm. For the probing microwave frequencies used, the filament geometrically corresponds to a one-third wavelength dipole antenna with a longitudinal depolarization factor of $\xi \approx 0.01$ [30]. Two conclusions can be drawn in the context of this work (based on plasma imaging and $n_e$-measurements detailed below). First, the depolarization factor appearing in the denominators of Eq. 4, Eq. 5, Eq. 6, Eq. 7, and Eq. 12 can be neglected restricting scattering to the collisional and Thomson regimes (see Methods for details). Second, the approximation of a short ($L < \lambda$) and thin ($D \ll \lambda, \delta$) plasma ellipsoid considered in the theoretical model above is satisfied in experiments (the 11 GHz skin depth condition allows for $n_e$ measurements up to $10^{20}$ m$^{-3}$ and $4 \cdot 10^{21}$ m$^{-3}$ for background pressures of 1 and 760 Torr, respectively, while actual measured $n_e$-values were lower as shown below).

We then scatter 10.5-12 GHz microwaves off the REMPI-induced microplasma – as schematically shown in Fig. 2(b), where the microwave electric field is oriented parallel to the plasma channel (see Methods for details). The radiation pattern measured by an I/Q mixer-based microwave system at t = 5 ns (after initiation of the laser pulse) is depicted in Fig. 2(c). Both the theoretical short-dipole (SD, line) and experimental (8 and 760 Torr, scatter) f = 10.5 GHz E-field amplitude ($\propto V_S$) radiation patterns at a far-field Tx/Rx distance of 10 cm are illustrated. It was observed that the dependence of the measured scattered signal on $\theta$ and $\phi$ closely resembles the short dipole for both the collisionless (8 Torr) and collisional (760 Torr) plasmas over the early timescales presented in this work. Specifically, the radiation patterns in Fig. 2(c) demonstrate a sinusoidal dependence with respect to $\theta$ and isotropic with respect to $\phi$. Further, the experimental pattern is indeed linearly polarized, as indicated by the $\beta$ plot.

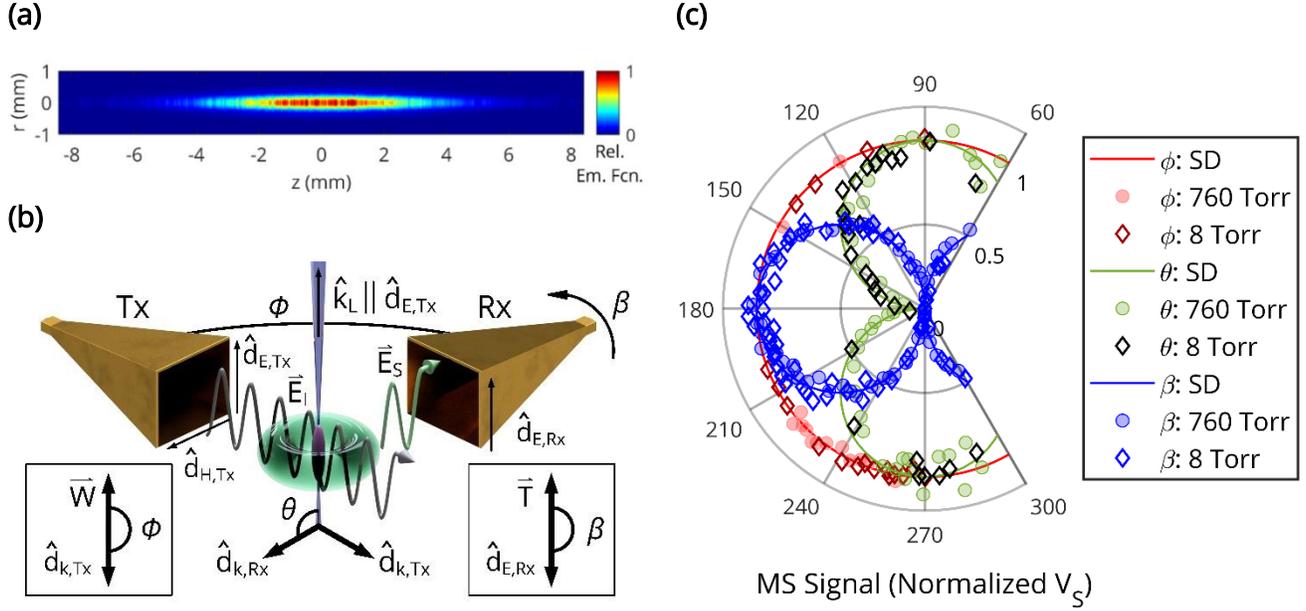

**Fig. 2 Coherent microwave scattering from short, thin weakly-ionized plasmas. (a)** Relative emission function (representative of an axisymmetric electron number density) of the laser-induced plasma at a background pressure of 760 Torr, calculated from an Abel-inversion of the corresponding ICCD image (gate width 100 ns, 250 accumulations). **(b)** Illustration of in-phase coherent microwave scattering and short-dipole-like far-field radiation behavior. **(c)** Short dipole (SD, line) and experimental f = 10.5 GHz plasma scattered (8 and 760 Torr, scatter) radiation patterns at t = 5 ns for a Tx/Rx distance of 10 cm. The microwave scattering signal (normalized Vs) is proportional to the scattered E-field amplitude, and angular variation is based on default positions $\phi = 270°$, $\theta = 90°$, and $\beta = 180°$.

Temporal evolution of total electron numbers $N_e$ in the plasma filament measured by 11 GHz microwave scattering at 1, 10, 100, and 760 Torr are shown in Fig. 3(a) – along with the normalized laser photodiode signal. The associated line-of-sight ICCD images at 100 and 760 Torr are additionally illustrated in Fig. 3(b). One can see that $N_e$ rapidly increases during the ionizing laser pulse and then decays shortly after (governed by dissociative recombination, electron attachment, and diffusion [35] [36] [27] [37]). Observed decay times (> 5-7 ns) are consistent with others' computational and experimental work [4] [36] [9] [7] [38].

The total number of electrons produced by the laser pulse $N_{e,laser}$ (approximated by the peak $N_e$-value) is displayed on a logarithmic scale for four discrete microwave frequencies in Fig. 3(c). It can be seen that $N_{e,laser}$ scales nearly linearly with pressure in the range 1-100 Torr and begins to saturate (less-steep slope) for 300-760 Torr. This saturation onset at higher pressures can be explained by electron losses during the laser pulse and, as a result, underestimation of the actual $N_{e,laser}$-value for $P \geq 300$ Torr. For example, an electron decay time of ~ 7 ns was observed at 760 Torr which is comparable with the laser pulse duration of ~ 5 ns. In contrast, observed electron decay times substantially exceeded the laser pulse duration for $P \leq 100$ Torr and, therefore, $N_{e,laser}$ can be reliably approximated by the peak $N_e$-value in that range.

Regions of Thomson, mixed, and collisional scattering regimes are indicated on Fig. 3(c) and correspond to the pressure ranges 0-10 Torr, 10-100 Torr, and 100-760 Torr, respectively. The boundaries of these regions can be estimated using the theoretical model above corresponding to the value of the effective collisional frequency for momentum transfer $\nu_m$ ($\nu_m \propto$ pressure) in relation to the microwave angular frequency $\omega$. The plasma dynamics measured at 760 Torr corresponds to the collisional scattering regime and is consistent with previous works. A maximum total electron number of $N_e = 3 \cdot 10^{11}$ is correspondingly observed and, based on the previous Abel-inverted upper bound on the volume, the peak spatially averaged electron number density can be estimated on the order $10^{20} \frac{1}{m^3}$. This density agrees well with similar studies in literature [7].

In this work, in-phase coherent scattering of microwaves off a microplasma at reduced gas pressures in the Thomson and mixed collisional-Thomson regimes is demonstrated for the first time. The Thomson regime is particularly advantageous in that a detailed knowledge of collisional properties is unnecessary to characterize scattering. That is, an accurate description of $\nu_m$ (often difficult to determine due to unknown local temperatures and species densities, inhomogeneities, and so on) is not required to accurately extract $N_e$ from the measured scattered signal $V_S$ since the following simple relation holds in the Thomson regime: $N_e = \frac{m\omega}{Ae^2} V_S$ (see Eq. 14). Additionally, the Thomson regime has higher total electron number sensitivities than the collisional regime due to larger individual electron scattering cross sections.

The Thomson regime is particularly advantageous for the evaluation of photoionization rates and cross-sections in comparison with collisional scattering at 1 atm used previously [4] [5] [9] [10]. In addition to $v_m$-independent measurement results in the Thomson regime (as discussed in the previous paragraph), it also enables measurements for an extended range of laser intensities by delaying onset of nonlinear optical phenomena to higher laser intensities. This refers to the fact that nonlinear optical phenomena including Kerr and plasma nonlinear terms in the refractive index reduce proportionally with a decrease in pressure ($n_2 \propto n_g \propto P$, $\omega_p^2 \propto n_e \propto n_g \propto P$).

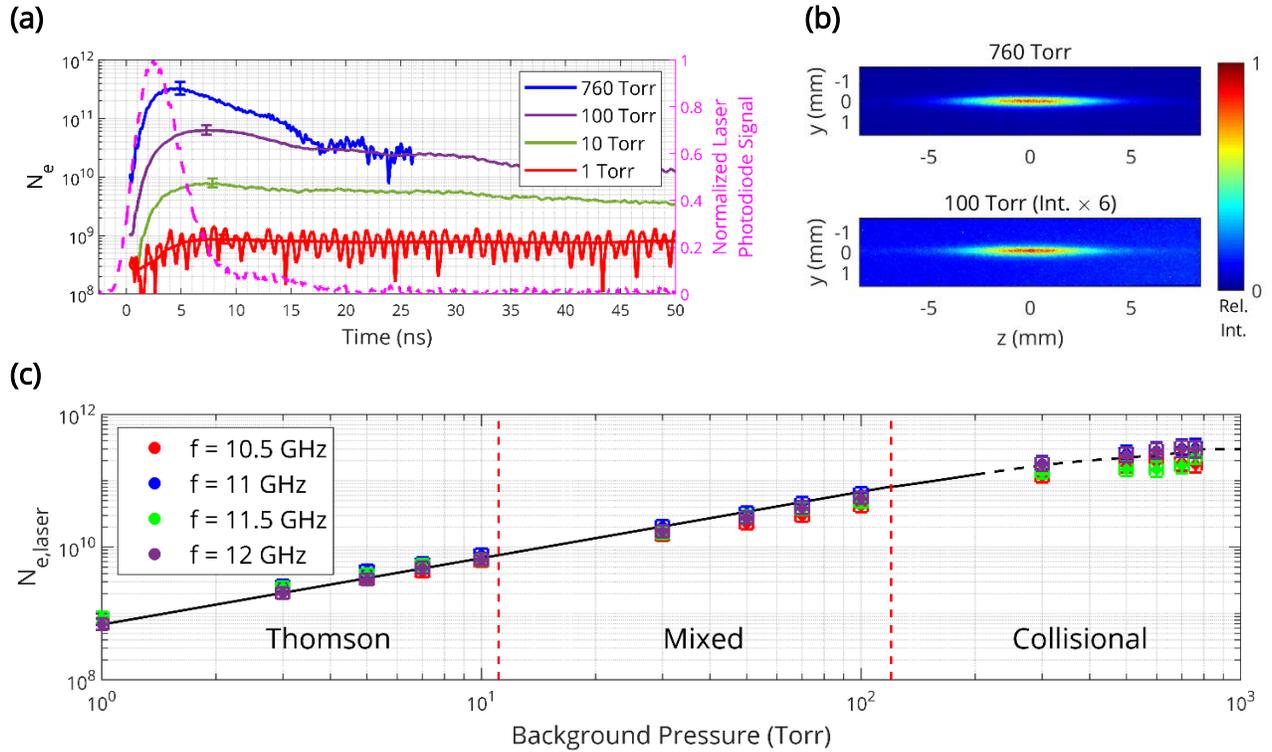

**Fig. 3 Coherent MS measurements of total electron numbers in variable pressure, short-dipole microplasmas assuming negligible depolarization and skin depth effects.** (**a**) Temporal evolution of the total electron number measured via 11 GHz microwave scattering with the dielectric scatterer calibration technique at 1, 10, 100, and 760 Torr (with error bars)– alongside (**b**) line-of-sight intensified charge-coupled device images (gate width 100 ns) of the corresponding laser-induced plasmas. (**c**) Total number of electrons produced by the laser (approximated by the peak $N_e$ value, which holds particularly well in the solid black-line region $P \leq 100$ Torr) as a function of background pressure for 10.5, 11, 11.5, and 12 GHz microwave scattering (with error bars).

Finally, we will experimentally verify the theoretical model of in-phase CMS by direct measurements of total scattering cross-sections of the dielectric calibration scatterer and measurements of the relative phase $\Delta\Phi$ (see Methodology for details). First, the total cross-section of the dielectric calibration sample was experimentally determined using direct measurements of the total power scattered by the sample and the incident microwave intensity at the bullet's location to calculate $\sigma_{Tot,D} = \frac{\langle P_S \rangle}{I_I}$. This experimental value was then compared with the one predicted by the theoretical model for the case of a Teflon dielectric scatterer $\sigma_{Tot,D} = \frac{\omega^2}{6\pi\varepsilon_0^2 c^4}(\varepsilon_0(\varepsilon_D - 1)\omega V_D)^2$, where $\varepsilon_D$ is the sample's dielectric constant, and $V_D$ is the sample's volume (electrical conductivity and depolarization factor of the bullet are assumed to be 0). Measurements matched the theoretical prediction within 20%. Second, relative measurements of the phase between the electron displacement and the incident electric field were conducted. In specific, the parameter $\Delta\Phi = \Phi - \Phi_{1\,Torr}$ (representing a relative phase which is set to be 0 at $P = 1$ Torr) was measured by the I/Q mixer-based microwave system in the pressure range 1-760 Torr. This was accomplished via measurement of the scattered radiation phase with respect to the local oscillator $\Phi_S$ through the $\bar{I}$ and $\bar{Q}$ mixer channels. $\Phi_S$ remained relatively invariant over the timescales presented in this work. The measured dependence of $\Delta\Phi$ (at $t = 3$ ns) on background pressure is depicted in Fig. 4 for 11 GHz CMS, alongside the theoretical relative phase of electron displacement from Eq. 5. One can see that the phase shift changes about 90 degrees during the transition from the Thomson to the collisional scattering regime. This is fundamentally rooted in the fact that $s$ and $E_I$ oscillate in-phase for free electrons in the Thomson regime ($\ddot{s} \propto -E_I$), while a 90-degree phase shift is introduced between $s$ and $E_I$ in the collisional regime ($\dot{s} \propto E_I$). Relatively good agreement between experimental data and theoretical predictions is additionally observed in Fig. 4.

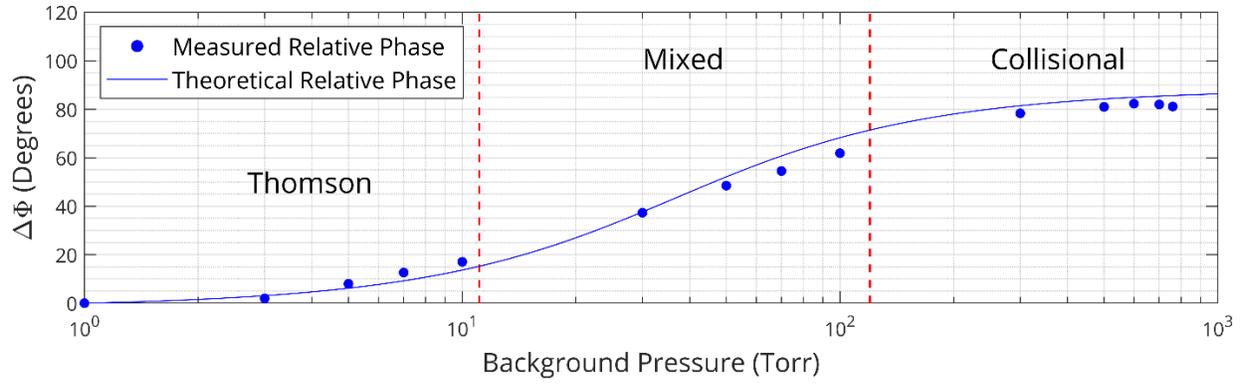

**Fig. 4 Relative phases for coherent microwave scattering as a function of background pressure.** The I/Q-mixer measured relative phase of scattered radiation $\Delta\Phi$ and theoretical prediction for 11 GHz microwave scattering at t = 3 ns.

# Methodology

A schematic of the general experimental setup is depicted in Fig. 5.

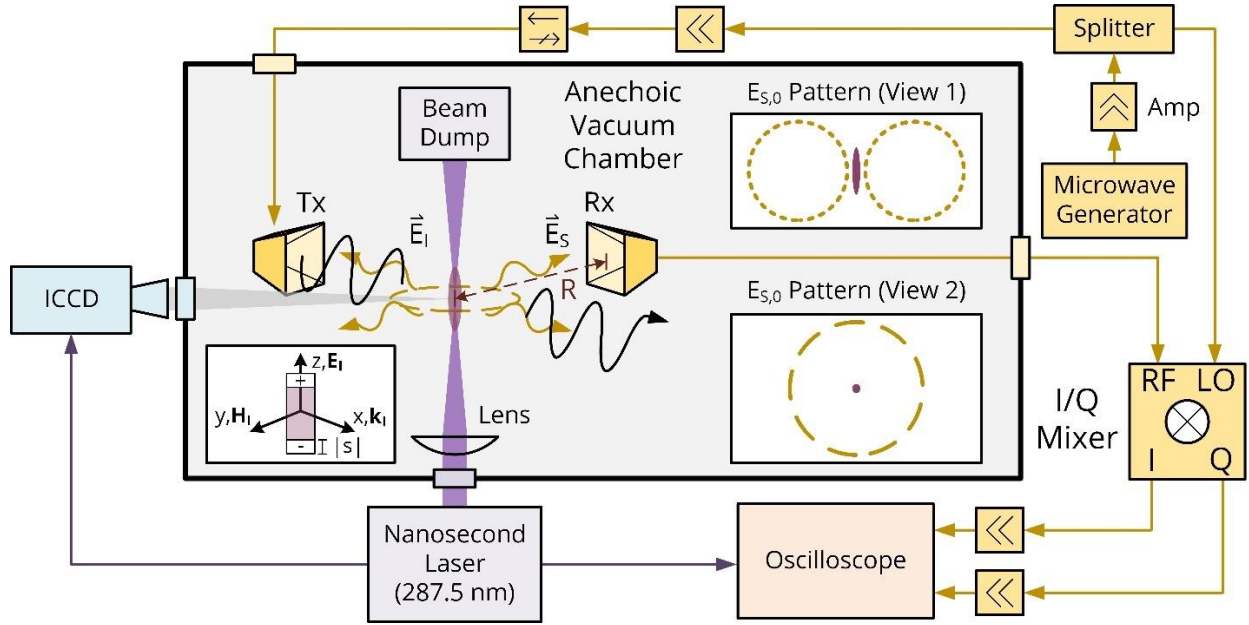

**Fig. 5 Experimental set-up.** Conceptual illustration of in-phase coherent microwave scattering in the far-field by a short, thin plasma object – coupled with a general schematic of the photoionization (purple), plasma imaging (blue), and microwave scattering (yellow) systems.

**Vacuum chamber.** Microwave scattering experiments were performed in a 0.66 m$^3$ vacuum chamber lined with carbon-coated polyurethane foam (insertion loss ~ 22 dB, reflectivity level < -18 dB). An MKS 925 Micro Pirani transducer (uncertainty 5%) was utilized for pressure monitoring. To ensure consistent air composition (P ≤ 760 Torr, T = 298 K, specific humidity ≤ 7 g/kg), the chamber was first mechanically pumped, purged and outgassed using a diffusion pump at 10$^{-5}$ Torr, sealed, then filled with ambient air for experiments. Minimization of outgassing was confirmed after purging through an MKS Microvision 2 residual gas analyzer. UV fused silica and N-BK7 windows were installed for laser delivery and plasma imaging, respectively. Hermetically sealed SMA feedthroughs were inserted for microwave line connections and compression vacuum tube fittings were utilized for external angular positioning of the Rx microwave horn.

**Laser-induced plasma generation.** 4.93 ± 0.25 ns FWHM laser pulses at 287.5 nm (FWHM 0.1 nm) were first produced by a 10 Hz, broadly tunable Ekspla NT342 Q-switched Nd:YAG laser system (1064 nm→ SHG → THG → OPO → OPO-SHG). The laser was measured to be nearly linearly-polarized with an axial ratio of > 10 – where the major axis is aligned parallel to the resultant of orthogonal unit vectors $\hat{d}_{H,Tx}$ and $\hat{d}_{k,Tx}$ (ref. Fig. 2). The 7 mm diameter top-hat beam with full divergence angle < 2 mrad was then focused through a f = 175 mm plano-convex spherical lens to a peak intensity on the order of $I_L \sim 10^9$ W/cm$^2$ for plasma generation (maximum pulse energy of 3 ± 0.38 mJ after the lens, measured with a pyroelectric energy meter). A silicon photodiode with a rise/fall time of < 1 ns was used for relative intensity measurements at the position of plasma (assuming a constant beam waist), and an electrical output signal from the laser (jitter 0.3 ns) was utilized for external device triggering. Pulse energy could be varied via Q-switch timing modifications. The laser was finally quenched by a beam dump positioned sufficiently-far from the focal point.

A notable ionization process is observed for diatomic oxygen at 287.5 nm through the (2+1) REMPI mechanism. Specifically, via 2-photon excitation $O_2\ C^3\Pi_g\ (v' = 2,\ J') \leftarrow O_2\ X^3\Sigma_g^-\ (v'' = 0,\ J'')$ followed by 1-photon absorption to ionization continuum $O_2^+$ [15]. The effective ionization cross section for (2+1) REMPI of $O_2$ is significantly greater than that for tunneling and non-resonant multiphoton ionization of air constituents [39] [40] [41] [42]. This feature was confirmed experimentally by significantly reduced ICCD and microwave scattering signals off resonance (282 and 290 nm). The resulting photoelectrons naturally exhibit angular and kinetic energy distribution signatures of the underlying REMPI process [40] [43]. Thermalization, reactions, and supplemental radiation heating quickly modify these distributions. For an order of magnitude estimate, the electron temperature can be evaluated through multiphoton excess $T_e \approx 3\hbar\omega_L - U_{I,O_2} = 0.868$ eV [8]. Supplemental electron heating is minimally expected for our laser conditions and multiphoton ionization can thus be interpreted as the primary contributor to electron temperature.

For the low energy depositions encountered in this work (non-detectable changes in laser pulse energy across the plasma on the precision 0.02 mJ) and nature of the REMPI mechanism, minimal optical perturbations to the neutral gas density and temperature are expected for the timescales presented [8] [38]. The plasma can be further regarded as predominantly weakly-ionized [8]. Despite REMPI-enabled selectivity, several species may arise over the laser pulse duration at sufficiently-high pressures due to secondary ionization (e.g. electron collisions with metastable species) and a myriad of reactions [35] [36] [37] [13] – a set of processes hastened by high transport frequencies. A nonequilibrium air-based, weakly-ionized plasma is consequently considered for decay rate comparisons.

**Plasma imaging.** Photographs of the photoionized plasmas were taken using a Princeton Instruments PI-MAX 4: 1024i-RB ICCD camera with a lens of focal length $f_{ICCD}$ = 55 mm (f stop $f_{ICCD}$/2.8). The camera was triggered using the laser electrical output pulse and surveilled with a gate monitoring signal. In reality, the signals observed can be highly intricate – exhibiting effects of straylight, laser-induced fluorescence, laser light scattering (elastic and inelastic), plasma emission, etc. [44] [45]. Straylight contributions were minimized through short gate times and a proper background environment, as validated through near-zero signals far-off the main laser line. The bulk of laser-induced fluorescence and laser light scattering processes were additionally found to be weak – as justified / confirmed through low N-BK7 transmissivity and camera quantum efficiencies in the ultraviolet range, imaging of off-resonant laser wavelengths (282 and 290 nm), and relative intensity comparisons with the corresponding microwave scattering signals for various photoionization conditions and ICCD gate features (delay and width). Some competition to plasma-related emission (see [46] for relevant processes) is expected from two-photon-excited fluorescence (TPEF). However, overlap and intensity scaling similarities between TPEF and the (2+1) REMPI process roughly preserve the perceived plasma size. It should be noted that emission does not explicitly indicate the presence of electrons but can provide a reliable order-of-magnitude estimate of the plasma size [9]. An improved method for an estimation would entail the isolation of electron-induced fluorescence (see studies on the second positive system of $N_2$ [47]). Further, the filament volume and distribution will be unique for a given pressure and timestamp due to optical effects [44] [48], diffusion, and diverse decay / reaction mechanisms [27].

In the case where emission is linearly proportional to the electron number density, a line-of-sight ICCD image represents the graphical projection of a three-dimensional plasma object rather than $n_e$. For improved insight on the electron number density distribution and plasma volume, an Abel inversion ($g(r,z) = -\frac{1}{\pi}\int_y^R \frac{dh(y,z)}{dy} \cdot \frac{dy}{\sqrt{y^2-r^2}}$, where g(r,z) is an emission function approximately proportional to $n_e$ and h(y,z) is the corresponding projection) was applied to the photograph in Fig. 2(a) under a reasonable assumption of axial symmetry. A Fourier-based algorithm was utilized for the inversion [49].

**Microwave scattering (MS) system.** The microwave field scattered off the plasma volume was measured by a receiving (Rx) horn coupled with an in-phase / quadrature (I/Q) mixer-based homodyne detection system that provides an output voltage $V_S \propto E_{S,0}$ (where $E_{S,0}$ is the amplitude of the scattered field at the Rx horn location) – as illustrated in Fig. 5 and described in detail below. First, an Anritsu 68369B synthesized signal generator was used to produce a continuous microwave signal at frequencies 10.5-12 GHz (FWHM 1 MHz). Such frequencies weakly interact with the neutral species considered in this work [50] [51]. The microwaves were then amplified and split. One arm was connected to a MITEQ IRM0812LC2Q I/Q mixer local oscillator (LO) port, with a confirmed operational power in the linear mixer regime (10 – 13 dBm). The other arm was then further amplified, isolated, and sent to a pyramidal 20 dB transmitting (Tx) horn for plasma irradiation. The linearly polarized microwave intensity at the plasma interface was measured to be ~ 3.5 W/m² (relatively low incident microwave intensities were utilized to minimize plasma perturbations [39] and nonlinear interactions [48]). Negligible heating by probing microwaves were supplementally confirmed though the invariance of plasma decay behavior on modest increases in supplied microwave power (0.1 W → 6 W). A 15 dB receiving horn was then installed and connected to the I/Q mixer radio frequency (RF) port to detect the scattered electric field. Both the transmitting and receiving horns were placed inside the vacuum chamber at a proper far-field distance of 10 cm from the plasma ellipsoid center - where sealed SMA feedthroughs were installed for external connections to the I/Q mixer. The in-phase ($\bar{I}$) and quadrature ($\bar{Q}$) channels of the mixer were subsequently connected to a WavePro 735Zi 3.5 GHz oscilloscope with 50 Ohm DC termination. Calibrated amplification of the MW scattering signals (before the RF port) and the I/Q channels were used as needed to improve system sensitivity (at the consequence of uncertainty) and linear, undistorted amplifier operation was ensured throughout experiments.

A schematic of the plasma-microwave scattering experiment is shown in Fig. 2(b). The corresponding notation is defined as follows: $R$ – distance from the plasma ellipsoid center to the Rx detector central point, $E_I$ – incident electric field at the location of plasma, $E_S$ – scattered electric field at the observation location, Tx – transmitting horn, Rx – receiving horn, $\hat{d}_{k,Tx/Rx}$ – horn unit vector parallel to the wave vector when operating as a transmitter, $\hat{d}_{E,Tx/Rx}$ – E-plane horn flare unit vector, $\hat{d}_{H,Tx/Rx}$ – H-plane horn flare unit vector. Three angles are further defined: $\theta$ is the polar angle between the unit dipole vector ($= \hat{d}_{E,Tx}$) and $-\hat{R}$ ($= \hat{d}_{k,Rx}$), $\beta$ is the angle between the Rx horn E-field flare vector $\hat{d}_{E,Rx}$ and $\left(\hat{R} \times \left(\hat{R} \times \hat{E}_I\right)\right) = \vec{T}$, and the azimuthal angle $\phi$ (in which $E_{S,0}$ is ideally invariant under rotations of $\phi$ for a short dipole) is additionally defined between $\hat{d}_{k,Tx}$ and the projection $\text{proj}_U(\hat{d}_{k,Rx}) = \vec{W}$ – where $U$ is the plane formed by vectors $\hat{d}_{k,Tx}$ and $\hat{d}_{H,Tx}$. Default positions $\phi = 270°$, $\theta = 90°$, and $\beta = 180°$ are referenced for the subsequent angular variation of the Rx horn in experiment.

The amplitude of the scattered electric field ($E_{S,0}$) can be related to the linear I/Q voltage expressions as follows:

$$V_{\bar{I}} = \frac{\kappa B_{LO}}{2} \eta B_B \cos(\Phi_B) + \frac{\kappa B_{LO}}{2} \eta B_S \cos(\Phi_S(t)) = V_{\bar{I},0} + \Delta V_{\bar{I}} \qquad \text{Eq. 9}$$

$$V_{\bar{Q}} = \frac{\kappa B_{LO}}{2} \eta B_B \sin(\Phi_B) + \frac{\kappa B_{LO}}{2} \eta B_S \sin(\Phi_S(t)) = V_{\bar{Q},0} + \Delta V_{\bar{Q}} \qquad \text{Eq. 10}$$

$$\sqrt{\Delta V_{\bar{I}}^2 + \Delta V_{\bar{Q}}^2} = V_S \propto B_S \propto E_{S,0} \qquad \text{Eq. 11}$$

Where $\kappa$ is the conversion factor, $B$ is the local oscillator ($B_{LO}$) / background ($B_B$) / scattered radiation ($B_S$) signal amplitude, $\eta$ is the MW circuit attenuation factor, and $\Phi_B/\Phi_S$ are corresponding phases of background/scattered signals relative to the local oscillator. $\Phi_S$ can be determined through Eq. 9 and Eq. 10 as $\Phi_S = \tan^{-1}(\Delta V_{\bar{Q}}/\Delta V_{\bar{I}})$. For an improved signal-to-noise ratio and trial consistency (mitigating variations in laser pulse energies, vibrations, etc.), averaging over 120 events was accomplished via low-jitter laser triggering of the oscilloscope. Microwave scattering data at various probing frequencies were collected for diverse laser-induced air-based plasmas with background pressures in the range of 1-760 Torr.

**Electron number measurements.** The total number of electrons in the produced small plasma objects were determined from $V_S$ measurements coupled with MS system calibration via dielectric scattering samples. For CMS, scaling of $V_S$ with $N_e$ can be generally derived based on results obtained in previous sections (see Eq. 6, Eq. 7, Eq. 8 and Eq. 11) as:

$$V_S \propto E_{S,0}|\cos(\beta)| = \sqrt{\frac{3\langle P_S \rangle}{4\pi\varepsilon_0 c}} \frac{|\sin(\theta)\cos(\beta)|}{R} = \frac{e^2\omega^2 E_{I,0}}{4\pi R m \varepsilon_0 c^2} \frac{|\sin(\theta)\cos(\beta)|}{\sqrt{\left(\xi\omega_p^2 - \omega^2\right)^2 + (\nu_m\omega)^2}} N_e \qquad \text{Eq. 12}$$

Where the total electron number measurements were conducted in the default configuration $\phi = 270°$, $\theta = 90°$, $\beta = 180°$, $R = 10$ cm as shown in Fig. 2(b). Highly prolate plasma ellipsoids ($\xi < 0.01$) with moderate-to-low electron number densities were considered such that the $\xi\omega_p^2$ term in the denominator of Eq. 12 is negligible and only the collisional or Thomson scattering regimes can be observed. In this case, scaling of $V_S$ with $N_e$ can be presented as $V_S \propto \frac{e^2 E_{I,0}}{4\pi R m \varepsilon_0 c^2} \frac{\omega}{\sqrt{\omega^2+\nu_m^2}} N_e$. One particularly convenient form of the $V_S - N_e$ relation (used in this work) is shown in Eq. 13, where $A$ denotes an overall calibration factor of the MS system.

The calibration factor $A$ was determined for each microwave frequency using a cylindrical PTFE sample with known relative permittivity $\varepsilon_D = 2.1$ and volume $V_D$ (length - 12.7 mm, diameter - 3.175 mm). The sample was mounted at the position of plasma, where the cylinder's axis was oriented parallel to the laser wave vector. Assuming the PTFE bullet to be sufficiently short and thin with minimal depolarization effects ($E_{D,0} \approx E_{I,0}$), the dependence of measured $V_S$ on dielectric scatterer properties can be written as the lower expression in Eq. 13 [3] [34]. This relation was utilized to determine the calibration factors $A$.

$$V_S = \begin{cases} A\dfrac{e^2}{m\sqrt{\omega^2+\nu_m^2}} N_e & \text{– plasma scatterer} \\ A V_D \varepsilon_0 (\varepsilon_D - 1)\omega & \text{– dielectric scatterer} \end{cases} \qquad \text{Eq. 13}$$

The total number of electrons $N_e$ in the produced plasma objects were finally evaluated using the following procedure. First, the calibration factor $A$ was determined for each microwave frequency from $V_S$ measurements of the dielectric scattering sample using the lower expression in Eq. 13. For example, the correspondence constant for 11 GHz scattering was measured to be $48.4 \frac{V\Omega}{cm^2}$. Then, the derived calibration factors $A$ and $V_S$ measurements of the plasma scatterers were used via the upper expression to determine $N_e$ values. The effective collisional frequency for momentum transfer $\nu_m$ was estimated as a function of pressure under the assumption of a weakly-ionized air plasma with minimal laser-perturbations to the neutral background, a constant electron temperature of $T_e = 0.9$ eV $= 10444$ K, and isotropic, elastic electron-neutral scattering. That is, $\nu_m = n_{O_2}\sigma_{eO_2}v_{Te} +$ $n_{N_2}\sigma_{eN_2}v_{Te} \approx 1.88 \cdot 10^9 \cdot P(Torr)$, where $\sigma_{eg}$ represents the electron-gas momentum transfer cross section ($\sigma_{eO_2} \approx 7.4 \cdot 10^{-20}\ m^2, \sigma_{eN_2} \approx 9.74 \cdot 10^{-20}\ m^2$ for $T_e = 0.9$ eV [52]), $v_{Te} = \sqrt{\frac{8k_bT_e}{\pi m}}$ is the thermal electron velocity, and $n_g$ is the neutral gas density. Despite often difficult to characterize dependencies on temperature, present species / reactions, specie densities, electron-ion collisions, etc. [27] [34], this estimate remains relatively reliable for the considered plasma and experimental conditions [5] [9] [53].

Note that Eq. 13 reduces to the following form when extreme cases of collisional ($\nu_m \gg \omega$) and Thomson ($\nu_m \gg \omega$) scattering are considered:

$$V_S = \begin{cases} A\frac{e^2}{m\omega}N_e & - \quad \text{plasma scatterer (Thomson regime)} \\ A\frac{e^2}{m\nu_m}N_e & - \quad \text{plasma scatterer (Collisional regime)} \\ AV_D\varepsilon_0(\varepsilon_D - 1)\omega & - \quad \text{dielectric scatterer} \end{cases} \qquad \text{Eq. 14}$$

Uncertainty in $N_e$ measurements are attributed to skin depth and collective scattering effects, nonidealities of the MW system, and nonlinear laser phenomena.

**Scattered radiation pattern measurements.** The microplasma-scattered radiation pattern in the far field was assessed at 8 and 760 Torr for 10.5 GHz microwaves. In these experiments, the MW signal was not amplified to prevent mixer saturation. Further, data was not collected in inaccessible locations or positions where an excessive RF-port power level was experienced. Angular variation is based on the default configuration $\phi = 270°$, $\theta = 90°$, $\beta = 180°$, $R = 10$ cm, as referenced in Fig. 2, where compression vacuum tube fittings were utilized for external angular positioning of the Rx horn. Data for the plots in Fig. 2(c) were generated by varying one angle at a time (namely, $\phi, \theta, \beta$) starting every time from the default configuration.

**Scattering cross section measurements.** The total scattering cross section $\sigma_{Tot} = \frac{\langle P_S \rangle}{I_I}$ was determined using radiation pattern data and measurements with the same MS detection system at two locations. First, $V_{S,def} \propto \sqrt{I_{S,def}}$ was measured in the default position ($\phi = 270°$, $\theta = 90°$, $\beta = 180°$, $R = 10$ cm). Using the experimentally confirmed scattered radiation pattern $I_S(\theta,\phi) = I_{S,def}\sin^2(\theta)$, the total time-averaged scattered power $\langle P_S \rangle$ can be derived as $\langle P_S \rangle = \oiint I_S(\theta,\phi)dS = \oiint I_{S,def}\sin^2(\theta)\,dS = \frac{8\pi}{3}R^2 I_{S,def}$. Next, the incident microwave intensity $V_I \propto \sqrt{I_I}$ was measured at the position of plasma (the Rx horn facing the Tx horn at $R = 0$, $\beta = 180°$) when the plasma-generating laser was disabled. The total scattering cross section $\sigma_{Tot}$ was then calculated from measured quantities $V_I$ and $V_{S,def}$ (note that $V_{S,def} \propto 1/R$):

$$\sigma_{Tot} = \frac{\langle P_S \rangle}{I_I} = \frac{\oiint I_S dS}{I_I} = \frac{I_{S,def}}{I_I}\frac{8\pi}{3}R^2 = \left(\frac{V_{S,def}}{V_I}\right)^2 \frac{8\pi}{3}R^2 \qquad \text{Eq. 15}$$

## Data Availability

The data that support the findings of this study are available from the corresponding author upon reasonable request.

## Acknowledgements

This work was supported by the National Science Foundation (Grant No. 1903415) and the Directorate for Mathematical and Physical Sciences (No. 1903360).

## Author Contributions

Experimental design, data collection, post-processing, and figure preparation were performed by A.R.P., A.R., and X.W. Theoretical models and the interpretation of results were composed by A.R.P., A.S., M.N.Sh., and M.N.Sl. All authors contributed to the manuscript. This collaborative effort was led and organized by A.S.